\documentclass{ws-ijmpcs}

\begin{document}

\markboth{Volker D. Burkert}
{Excited Nucleons and their Structure}

%
%

\title{STUDY OF EXCITED NUCLEONS AND THEIR STRUCTURE 
}
\author{VOLKER D. BURKERT}

\address{Jefferson Lab\\12000 Jefferson Avenue, Newport News, Virginia 23606, USA\\
burkert@jlab.org}
\date\currenttime
\maketitle
\begin{abstract}
Recent advances in the study of excited nucleons are discussed. 
Much of the progress  has been achieved 
due to the availability of high precision meson production data
in the photoproduction and electroproduction sectors, 
the development of multi-channel partial wave analysis techniques, 
and advances in Lattice QCD with predictions of the full excitation spectrum.     
\keywords{Nucleon resonances; Meson photoproduction; Helicity amplitudes.}
\end{abstract}
\ccode{PACS numbers: 14.20.Gk, 25.30.Rw, 25.20.Lj}

\section{Introduction}	

The $N^*$ series of workshop focuses on the part of the nucleon 
landscape where the excited states are prominent 
features in cross sections and polarization observables. Clearly, we 
can understand the complexity of the nucleon's internal structure 
only if we also study the excitation spectrum and the underlying degrees
of freedom. In this first talk I highlight 
some of the progress the community has made over the past 
five years or so, and touch on some of the very recent developments  that
we will discuss in course of this conference. 
The study of the $N^*$ structure (I use $N^*$ here generically for the 
non-strange spectrum of isospin $1\over 2$ nucleon and isospin ${3\over 2}$ 
$\Delta$) is not only  an important subject of hadron physics, but it is fundamental
to our understanding of the origin of the hadronic mass scale. 

One of the experimental tools we employ is the study of the $N^*$ spectrum 
as a reflection of the underlying degrees of freedom. Excited states could 
be represented by a bag of 3 constituent quarks, possibly held together
by glue strings, that themselves can be excited to generate gluonic or 
hybrid baryons. Quarks may also cluster dynamically into quark-diquark
configurations with a different excitation spectrum, and resonances may be 
excited dynamically through meson-baryon interactions. These 
configurations can result in variations of the excitation spectrum which we
study with hadronic and electromagnetic probes. The second line of 
research is to measure the strength of the resonance excitations at
different distance scales. Beyond these experimental aspects there
has been significant progress in the theory sector, e.g. predictions 
from Lattice QCD, Dyson Schwinger equations of QCD, holographic QCD, 
dynamical approaches and others. A number of  
reviews\cite{Crede:2013kia,Aznauryan:2012ba,Aznauryan:2011qj,Tiator:2011pw,Klempt:2009pi,Burkert:2004sk}
discuss many of these recent developments.   
\begin{figure}[t]
\vspace{-1cm}
\centerline{\psfig{file=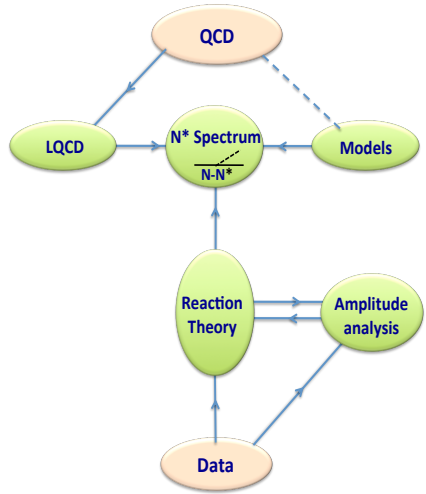,width=10.0cm,height=9cm}}
\caption{Schematic illustration of the interaction of experimental,  
phenomenological and theoretical aspects in the $N^*$ sector. 
 \label{analysis}}
\end{figure}
\section{Search for new Baryon States}
The strong hadronic couplings make most excited nucleon states very 
broad, often hundreds of MeV in width. Resonance parameters must 
be measured by measuring the coupling to the hadronic final states. This 
involves a separation of resonance and non-resonant processes that can both
lead to the same final state.  As a consequence resonances of different 
masses overlap and interfere. The search for new baryon states  
requires large amounts of data on cross section and polarization 
observables.  
Most of the states with significant coupling to single pion production may 
have been found already in elastic $\pi N \to \pi N$ scattering. For the 
past decade the focus has shifted to photoproduction. The search 
for new states is aimed at high statistics data sets with complete or nearly 
complete kinematic coverage of a number of two-body final states, such as 
$\gamma p \to \pi N,~\eta p,~KY$, both on proton and on neutron
 targets.  Some of the higher mass states may couple to
 more complex final states such as $\gamma p \to p \omega,~p \phi,~K^*Y$ or 
 multi-meson channels, e.g. $\gamma p \to \pi\pi N, \eta\pi N$. 

Essential component for the success of the effort 
is the engagement of groups involved in  
single channel coupled-channel analyses. Figure~\ref{analysis} illustrates
the interaction of  experimental data, phenomenological models and theory
predictions. QCD through lattice or  DSE, or model approximations make 
predictions for the spectrum and e.m. couplings. The data are input to amplitude 
analysis supported by reaction theory to extract the spectrum and transition 
amplitudes. 
\subsection{Complete experiments}
\begin{figure}[mh]
\centerline{\psfig{file=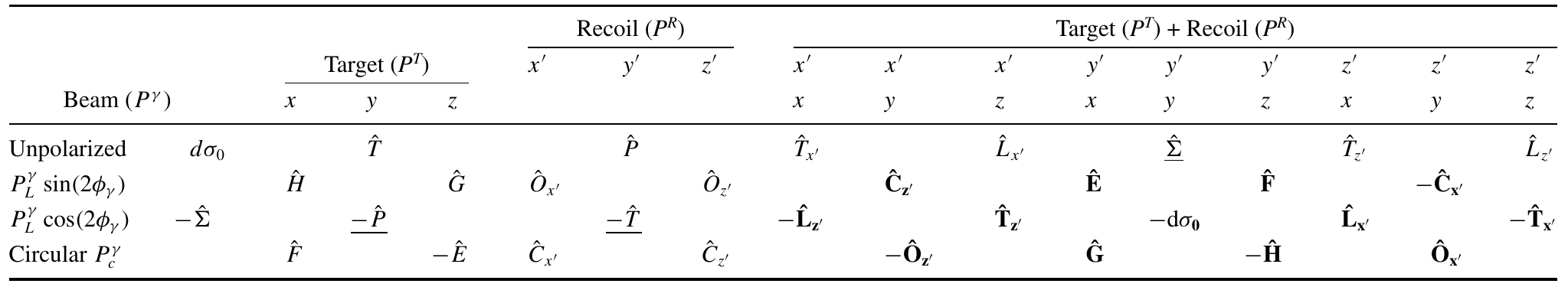,width=12.5cm,height=3cm}}
\caption{The 32 observables that can be accessed in single pseudoscalar  meson 
photoproduction using linear and circular polarized photons, longitudinal 
and transverse polarized targets, and recoil polarization.\protect\cite{Sandorfi:2010uv}
\label{complete}}
\end{figure}

The experimental effort has shifted from the nearly exclusive use of 
hadronic probes to electromagnetic probes. In the search for new 
baryon states, complete or even over-complete experiments have 
become the holy grail of baryon resonance analysis. The photo-production of 
single pseudo-scalar mesons is described by 4 complex parity 
conserving amplitudes requiring 8 well-chosen measurements to 
determine the production amplitudes. The observables are shown 
in Fig.\ref{complete}. In unpolarized, single and double polarization 
experiments, 16 observables can be measured directly. Of the 16 observables 3 
double polarization observables measure also single polarization quantities
(underlined in Fig.~\ref{complete}). Additionally, 13 triple polarization quantities 
are related to the cross section and 12 double polarization observables (bold face). 
While the 32 measurements are not independent, knowledge of as many polarization
quantities as possible helps to tighten constraints\cite{Sandorfi:2010uv} 
and allows for systematic cross checks. 
Some double polarization observables have recently been published\cite{Thiel:2012yj}.
\subsection{Open strangeness photoproduction}  
\begin{figure}[tpb]
\vspace{-0.5cm}
\centerline{\psfig{file=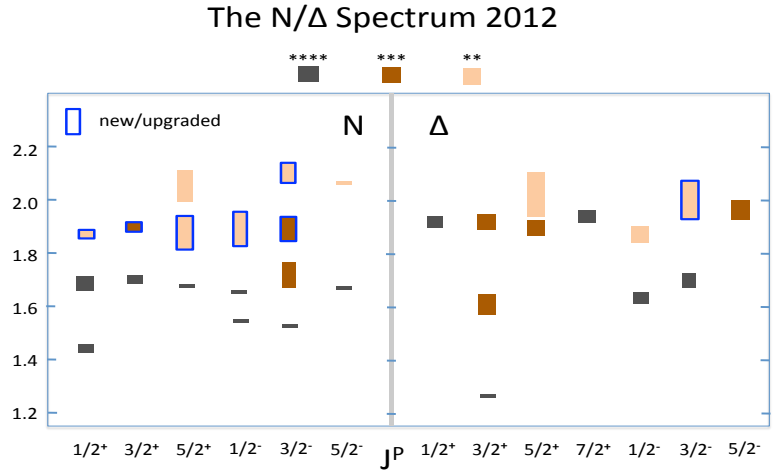,width=12.0cm,height=8cm}}
\vspace*{-1cm}
\caption{Nucleon and $\Delta$ states showing the changes 
in the 2012 RPP edition. The frames indicate states that are either new or have 
higher star rating compared to previous editions.
 \label{spectrum2012}}
\end{figure}

For the first time strangeness photoproduction has played a major 
role in the search for new baryon states. The very precise cross section 
and polarization data from CLAS\cite{McCracken:2009ra,Dey:2010hh,Bradford:2006ba,Bradford:2005pt} 
in the channels $\gamma p \to K^+ \Lambda, K^+ \Sigma^\circ$, and 
$\vec{\gamma} p \to K^+ \vec{\Lambda}$, were critical  in providing evidence for 
new states and for improving evidence for poorly known states especially in 
the positive parity nucleon 
sector in the most recent coupled-channel analysis of the 
Bonn-Gatchina group\cite{Anisovich:2011fc}. Figure~\ref{spectrum2012} 
shows the changes in the excited nucleon and $\Delta$ spectrum that 
were included largely due to the inclusion of the strangeness photoproduction
data in the analysis.  
In this regard some comments on the $N(1900){3\over 2}^+$ may be appropriate.  
The state is clearly seen as a peak in the fully integrated $\gamma p \to K^+\Lambda$ 
cross section. It has been first identified as a $J^P={3\over 2}^+$ state in a $K\Lambda$ 
coupled-channel analysis\cite{Shklyar:2005xg}. Since then it has been more firmly
established in the coupled-channel analysis\cite{Anisovich:2011fc} with high reliability
and upgraded to a *** state in the 2012 RPP edition. Subsequently the state was 
confirmed\cite{Mart:2012fa,Maxwell:2012tn} in two single channel analyses 
using an effective Lagrangian approach. Moreover,
the multi-channel partial-wave analysis of Shrestha and Manley\cite{Shrestha:2012ep}   
finds also strong coupling of the state to the $K\Lambda$ channel. Given that 
the state is clearly seen in the total $K\Lambda$ photoproduction cross section, and has
during the last year consistently been identified as a $J^P={3\over 2}^+$ state in four independent 
analyses with a Breit-Wigner mass of $\approx 1900$MeV, the state may have passed the 
criteria of a four star state.     
 \subsection{New nucleon candidates from charmonium decays}
 A different approach in the search for N* states comes from BESIII 
 with studies of the decay $\psi^\prime \to p\bar{p}\pi^\circ$.  The $p\pi^\circ$ mass is 
 analyzed\cite{Ablikim:2012zk} and shows some of the well-known isospin 
 $1\over 2$ states. Above 2~GeV 
 a large, isolated enhancement was found to represent two new N* candidates
 at 2300 and 2570 MeV. An interesting aspect of this reaction is that it not only 
 selects isospin $1\over 2$ states but suppresses high spin states due to the 
 short range interaction involved in the $c\bar{c}$ annihilation that generates 
 the $N^*\bar{p}$ system. The suppression of higher spin states greatly 
 simplifies partial wave analysis. 
 \subsection{Vector meson photoproduction}
 Vector meson production, e.g. $\gamma p \to p\omega$ and $p \phi$ has 
 generally not been incorporated in multi-channel analyses. Single channel analyses
 of high precision data on $p\omega$ have revealed sensitivity to several excited nucleon 
 states\cite{Williams:2009aa,Williams:2009ab} and show evidence for 
 "missing" states, such as the 2-star $N(2000){5\over 2}^+$. The role of the $N(1535){1\over 2}^-$
 has been studied in $\pi N \to N\phi$ \cite{Doring:2008sv} and new high precision 
 data on $\gamma p \to p\phi$ have become available\cite{Seraydaryan:2013}, showing 
 features that may indicate the presence of excited states. To explore these 
 possibilities, these data must be included in coupled-channel analyses.
  \section{Structure of excited nucleons and hybrid baryons}
 Electron scattering off nucleons probes the internal structure and the effective
 degrees of freedom at distance scales given by the inverse of the 
 momentum transfer $1/|Q|$ to the $NN^*$ system.  
  Precise electroproduction data in channels such as 
 $ep\to e(p\pi^\circ, \pi^+ n,~p\eta,~p\pi^+\pi-$) have been collected over the past decade
 and have enabled extraction of the resonance electrocoupling\cite{Joo:2001tw,Ungaro:2006df}
 amplitudes in a wide range of $Q^2$. Precise
 amplitudes for the transition from the proton ground state to the 
 $\Delta(1232){3\over 2}^+$, $N(1440){1\over 2}^+$, $N(1520){3\over 2}^-$ and 
 $N(1535){1\over 2}^-$ excited states, have been determined 
 using Unitary Isobar Model and Dispersion Relations\cite{Aznauryan:2011qj,Tiator:2011pw,Aznauryan:2009mx}, and reveal strong  meson-baryon contributions at low $Q^2$ for several 
 of the states, and 3-quark dominance at higher $Q^2$. 
 Further transition towards elementary quark-gluon degrees of
 freedom is expected to occur at $Q^2 \ge 6$GeV$^2$, which is reachable only at the 
 Jefferson Lab upgrade to 12GeV beam energy.
 \begin{figure}[b]
\centerline{\psfig{file=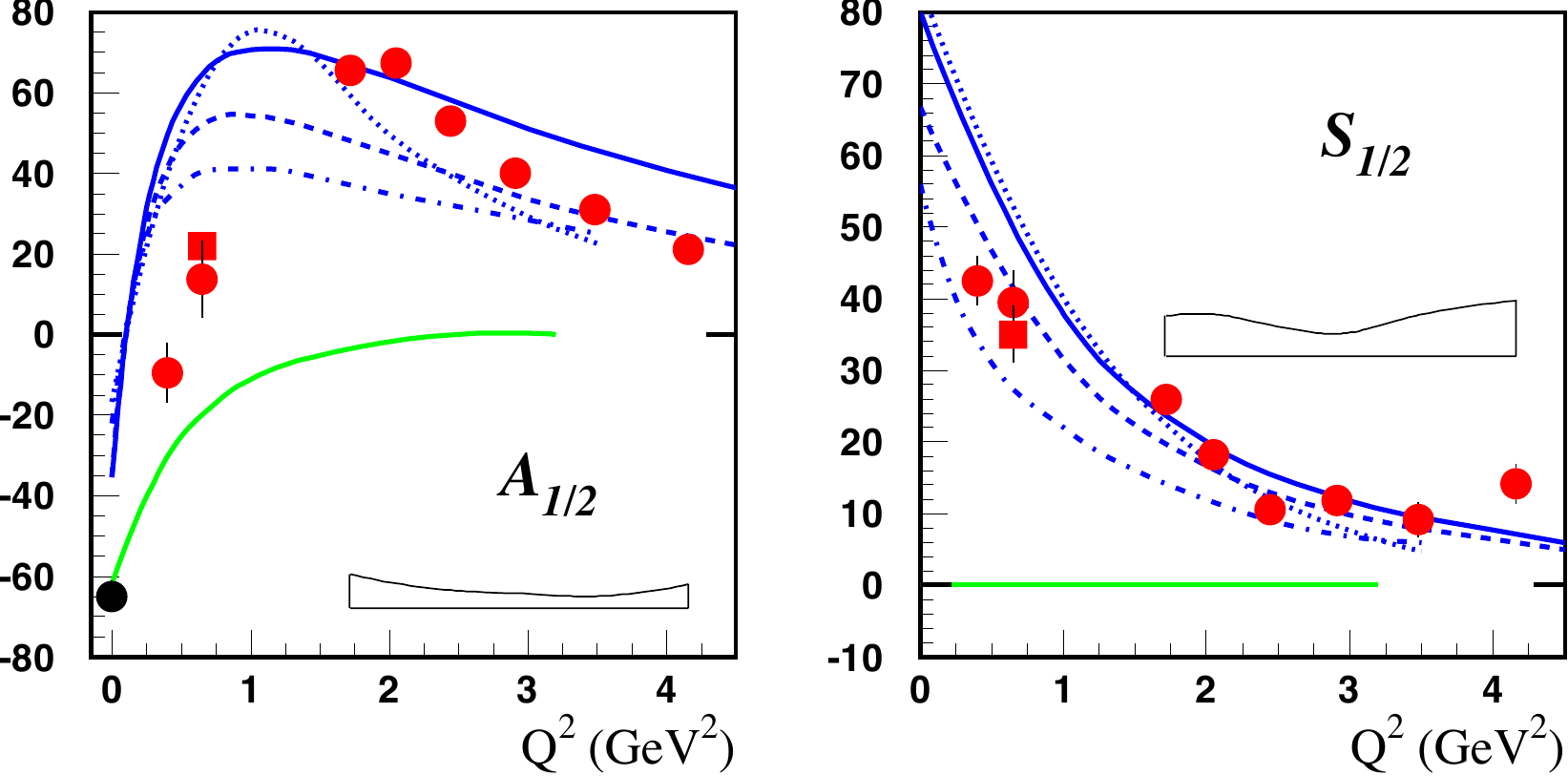,width=10.0cm,height=5.0cm}}
\caption{Helicity amplitudes for the Roper $N(1440)1/2^+$ state\protect \cite{Aznauryan:2008pe},
The lines close to the data are light cone quark models predictions. The 
solid green lines below are are predictions of a hybrid quark model. The $A_{1/2}$ 
amplitude drops quickly to 0, and the longitudinal amplitude $S_{1/2}=0$ in leading order. 
 \label{hybridRoper}}
\end{figure}

Probing the nucleon excitation with electron beams may be the only way of identifying 
hybrid baryons as they are not distinct by quantum numbers from ordinary baryons,
 and thus can mix with the quark states.    
In a simple quark model with gluon degrees of freedom\cite{Li:1991yba}, the lowest hybrid state 
corresponds to the transition in the quark ground state without orbital excitation leading 
to $J^P={1\over 2}^+$. The form of the $\gamma q \to \gamma G$ vertex makes the transition 
form factor drop rapidly with $Q^2$. In the leading term the longitudinal component is 
exactly zero. Figure~\ref{hybridRoper}\cite{Aznauryan:2008pe} shows data of the 
lowest ${1\over 2}^+$ state. 
Clearly,  the data have a completely different behavior, both for the transverse and the 
longitudinal amplitude, allowing us to conclude that the lowest ${1\over 2}^+$ state is 
not a hybrid baryon. 
The latest LQCD projections confirm the conclusion that this state is not the lowest hybrid 
baryon\cite{Dudek:2012ag}, which supports the conclusion from the transition form factor 
measurements as a probe of the nature of the states and the underlying degrees of freedom. 

\section{Outlook}  
During the past few years a major milestone was reached with experiments collecting 
data that represent 
complete or nearly complete measurements, especially in open strangeness production. 
In the phenomenological sector extensive efforts are underway to cope with
the quantity and high quality of the new data in a theoretically satisfying way. At this conference, some 
of the fruits resulting from this effort will be discussed. The 
coming years promise to be a period of discoveries of new excited states and the confirmation 
or dismissal of poorly known states, that will lead 
us to a much better understanding of the systematics and the nature of the nucleon excitations.
It is an exciting time to be part of this effort. 

\section*{Acknowledgements}
This work was carried out under DOE contract No. DE-AC05-06OR23177.

\end{document}